\documentclass[12pt]{article}
\usepackage{amsmath,amsthm,amsfonts}
\usepackage{fullpage}
\usepackage{graphicx}

\newtheorem{theorem}{Theorem}

\title{The state complexity of $L^2$ and $L^k$}
\author{Narad Rampersad \\
School of Computer Science \\
University of Waterloo \\
Waterloo, ON, N2L 3G1 \\
CANADA \\
{\tt nrampersad@math.uwaterloo.ca}}

\begin{document}
\date{\today}
\maketitle

\begin{abstract}
We show that if $M$ is a DFA with $n$ states over an arbitrary alphabet
and $L=L(M)$, then the worst-case state complexity of $L^2$ is
$n2^n - 2^{n-1}$. If, however, $M$ is a DFA over a unary alphabet,
then the worst-case state complexity of $L^k$ is $kn-k+1$ for all $k \geq 2$.
\end{abstract}

\section{Introduction}
We are often interested in quantifying the complexity of a regular
language $L$.  One natural complexity measure for regular
languages is the \emph{state complexity} of $L$, that is, the number of
states in the minimal deterministic finite automation (DFA) that
accepts $L$.  Given an operation on regular languages, we may also define
the state complexity of that operation to be the number of states that
are both sufficient and necessary in the worst-case for a DFA to
accept the resulting language.

Birget \cite{Bir92} gave exact results for the state complexities
of the intersection and union operations on regular languages.
Yu, Zhuang, and Salomaa \cite{YZS94} studied other operations, such as
concatenation and Kleene star.  For instance, Yu, Zhuang, and Salomaa
proved that, given DFAs $M_1$ and $M_2$ with $m$ and $n$ states respectively,
there exists a DFA with $m2^n - 2^{n-1}$ states that accepts $L(M_1)L(M_2)$.
Moreover, there exist $M_1$ and $M_2$ for which this bound is optimal.
Some more recent work on the state complexity of concatenation has been
done by Jir\'askov\'a \cite{Jir05} as well as Jir\'asek, Jir\'askov\'a,
and Szabari \cite{JJS04}.  Birget's work \cite{Bir96} on the state
complexity of $\overline{\Sigma^*\overline{L}}$ may also be of interest.

We are interested here in the state complexity of the concatenation
of a regular language $L$ with itself, which we denote $L^2$.  We
show that the bounds of Yu, Zhuang, and Salomaa for concatenation
are also optimal for $L^2$.  In other words, if $M$ is a DFA with
$n$ states and $L=L(M)$, then the worst-case state complexity of
$L^2$ is $n2^n - 2^{n-1}$.  This bound, however, does not hold if
we restrict ourselves to unary languages.  Specifically, we show
that if $M$ is a DFA over a unary alphabet, then the worst-case
state complexity of $L^k$ is $kn-k+1$ for all $k \geq 2$.

We first recall some basic definitions.  For further details
see \cite{HU79}.  A \emph{deterministic finite automaton} $M$ is
a quintuple $M = (Q,\Sigma,\delta,q_0,F)$, where $Q$ is a finite
set of states; $\Sigma$ is a finite alphabet;
$\delta\,:\,Q \times \Sigma \rightarrow Q$ is the transition function,
which we extend to $Q \times \Sigma^*$ in the natural way; $q_0 \in Q$
is the start state; and $F \subseteq Q$ is the set of final states.
A DFA $M$ accepts a word $w \in \Sigma^*$ if $\delta(q_0,w) \in F$.
The language accepted by $M$ is the set of all $w \in \Sigma^*$ such that
$\delta(q_0,w) \in F$; this language is denoted $L(M)$.  We denote the
language $L(M)L(M)$ by $L^2(M)$.  We may extend this notation to higher
powers by the recursive definition $L^k(M) = L^{k-1}(M)L(M)$ for $k \geq 2$.

\section{State complexity of $L^2$ for binary alphabets}
In this section we consider the state complexity of $L^2$ for
languages $L$ over an alphabet of size at least 2.

\begin{theorem}
\label{main}
For any integer $n \geq 3$, there exists a DFA $M$ with $n$
states such that the minimal DFA accepting the language $L^2(M)$
has $n2^n - 2^{n-1}$ states.
\end{theorem}

\begin{proof}
That the minimal DFA for $L^2(M)$ has at most $n2^n - 2^{n-1}$
states follows from the upper bound of Yu, Zhuang, and Salomaa
for concatenation of regular languages mentioned in the introduction.
To show that $n2^n - 2^{n-1}$ states are
also necessary in the worst case we define a DFA
$M = (Q,\Sigma,\delta,0,F)$ (Figure~\ref{dfa_M}), where
$Q = \lbrace 0,\ldots,n-1 \rbrace$, $\Sigma = \lbrace 0,1 \rbrace$,
$F = \lbrace n-1 \rbrace$, and for any $i$, $0 \leq i \leq n-1$,
\begin{displaymath}
\delta(i,a)=
\begin{cases}
0 & \text{if $a=0$ and $i=1$,}\\
i & \text{if $a=0$ and $i \neq 1$,}\\
i+1 \bmod n & \text{if $a=1$.}
\end{cases}
\end{displaymath}

\begin{figure}[ht]
\centering
\includegraphics{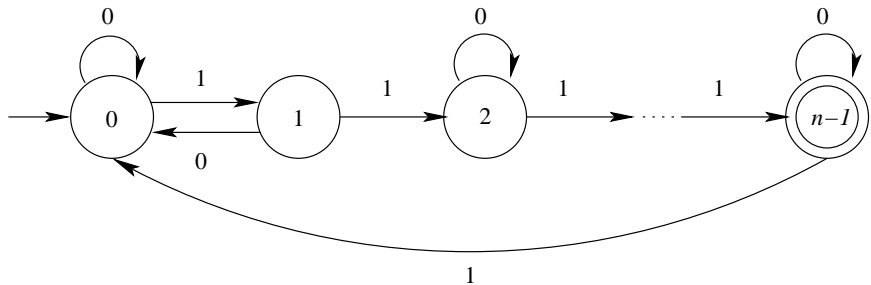}
\caption{The DFA $M$}
\label{dfa_M}
\end{figure}

We will apply the construction of Yu, Zhuang, and Salomaa
\cite[Theorem 2.3]{YZS94} and show that the resulting DFA for $L^2(M)$ is
minimal (see \cite{JJS04} for another example of this approach).
Let $M' = (Q',\Sigma,\delta',(0,\emptyset),F')$, where
\begin{itemize}
\item $Q' = Q \times 2^Q - F \times 2^{Q - \lbrace 0 \rbrace}$;
\item $F' = \lbrace (i,R) \in Q' \;|\; R \cap F \neq \emptyset \rbrace$;
and
\item $\delta'((i,R),a) = (\delta(i,a),R')$, for all $a\in\Sigma$, where
\begin{displaymath}
R'=
\begin{cases}
\delta(R,a) \cup \lbrace 0 \rbrace & \text{if $\delta(i,a) \in F$,}\\
\delta(R,a) & \text{otherwise.}
\end{cases}
\end{displaymath}
\end{itemize}
Then $L(M') = L^2(M)$ and $M'$ has $n2^n - 2^{n-1}$ states.

To show that $M'$ is minimal we will show (a) that all states of $M'$
are reachable, and (b) that the states of $M'$ are pairwise
inequivalent with respect to the Myhill--Nerode equivalence
relation \cite{Ner58,RS59}.

To prove part (a) let $(i,R)$ be a state of $M'$, where
$R = \lbrace r_1,\ldots,r_k \rbrace$.  If $0 \in R$, assume that
$r_k = 0$ and $r_1 < \cdots < r_{k-1}$; otherwise,
assume that $r_1 < \cdots < r_k$.  For $j = 1,\ldots,k$, define $s_j$
as follows:
\begin{displaymath}
s_j=
\begin{cases}
(r_j - 1) \bmod n & \text{if $j=1$,}\\
(r_j - r_{j-1}) \bmod n & \text{otherwise.}
\end{cases}
\end{displaymath}

If $i=0$, we see that \[\delta'((0,\emptyset),1^n(10)^{s_k} 1^n(10)^{s_{k-1}}
\cdots 1^n(10)^{s_1}) = (0,R).\]  If $i>0$, then let
$R' = \lbrace (r_1 - i) \bmod n, \ldots, (r_k - i) \bmod n \rbrace$.
Just as for $(0,R)$, we see that $(0,R')$ is reachable.  Moreover,
if $i \in F$ then $0 \in R$ and $1 \in R'$.  Hence,
$\delta'((0,R'),1^i) = (i,R)$, as required.

To prove part (b) let $(i,R)$ and $(j,S)$ be distinct states of $M'$.
We have two cases.

Case 1: $R \neq S$.  Then there exists $r$ such that $r$ is
in one of $R$ or $S$ (say $R$) but not both.  If $i \in F$, then
$r \neq 0$.  Hence $\delta'((i,R),1^{n-1-r}) \in F'$
but $\delta'((j,S),1^{n-1-r}) \not\in F'$.

Case 2: $R = S$.  Suppose $i<j$.  Let $i' = n-j+i$.  For
$T \subseteq Q$, let $T_{1 \rightarrow 0}$ denote the set
$(T \setminus \{1\}) \cup \{0\}$.  We have two subcases.

Case 2i: $((j+1) \bmod n) \not\in R$.  Then $\delta'((i,R),1^{n-j}) = (i',R')$
for some $R'$, and $\delta'((j,S),1^{n-j}) = (0,S')$ for some $S'$, where
$1 \not\in R'$ and $1 \in S'$.  We may now apply the argument of Case~1
to the states $(i',R')$ and $(0,S')$.

Case 2ii: $((j+1) \bmod n) \in R$.  If $i' \neq 1$, then
$\delta'((i,R),1^{n-j}) = (i',R')$ for some $R'$,
$\delta'((i',R'),0) = (i',R'_{1 \rightarrow 0})$, and
$\delta'((i',R'_{1 \rightarrow 0}),1^j) = (i,R'')$ for some
$R''$, where $((j+1) \bmod n) \not\in R''$.  Similarly,
$\delta'((j,S),1^{n-j}) = (0,S')$ for some $S'$,
$\delta'((0,S'),0) = (0,S'_{1 \rightarrow 0})$, and
$\delta'((0,S'_{1 \rightarrow 0}),1^j) = (j,S'')$ for some
$S''$, where $((j+1) \bmod n) \not\in S''$.
If $R'' \neq S''$, we apply the argument of Case~1 to the states
$(i,R'')$ and $(j,S'')$; otherwise, we apply the argument of Case~2i.

If $i'=1$, then since $i<j$, $i=0$ and $j=n-1$.  We thus have
$\delta'((0,R),0) = (0,R_{1 \rightarrow 0})$, and
$\delta'((0,R_{1 \rightarrow 0}),1) = (1,R')$ for some
$R'$, where $2 \not\in R'$.  Similarly,
$\delta'((n-1,S),0) = (n-1,S_{1 \rightarrow 0})$, and
$\delta'((n-1,S_{1 \rightarrow 0}),1) = (0,S')$ for some
$S'$, where $2 \not\in S'$.
If $R' \neq S'$, we apply the argument of Case~1 to the states
$(0,S')$ and $(1,R')$; otherwise, we apply the argument of Case~2i.
\end{proof}

\section{State complexity of $L^k$ for unary alphabets}
In this section we show that the bound given in Theorem~\ref{main}
does not hold if we restrict ourselves to unary languages.  We also
give optimal bounds for the state complexity of arbitrary powers
$L^k$ of a regular language $L$.

It is easy to see that the transition graph of a connected unary
DFA $M$ with $n$ states is composed of a ``tail'' with
$\mu \geq 0$ states and a ``cycle'' with $\lambda \geq 1$ states,
where $n = \mu + \lambda$.  Following Chrobak \cite{Chr86}, we
therefore denote the size of $M$ by the pair $(\lambda,\mu)$.

Pighizzini and Shallit \cite{PS02} give the following result regarding
concatenation of unary DFAs.

\begin{theorem}[Pighizzini and Shallit]
\label{pig_sha}
Let $L_1,L_2$ be unary languages accepted by DFAs of sizes
$(\lambda_1,\mu_1),(\lambda_2,\mu_2)$ respectively.  Then
there exists a DFA $M$ of size $(\lambda,\mu)$, where
$\lambda = \mathrm{lcm}(\lambda_1,\lambda_2)$ and
$\mu = \mu_1 + \mu_2 + \mathrm{lcm}(\lambda_1,\lambda_2) - 1$,
such that $L(M) = L_1L_2$.
\end{theorem}

From Theorem~\ref{pig_sha} we can derive the following
upper bound for the state complexity of $L^k$.

\begin{theorem}
Let $L$ be a unary language accepted by a DFA with $n$ states.
For all $k \geq 2$, there exists a DFA $M$ with $kn-k+1$ states
such that $L(M) = L^k$.
\end{theorem}

\begin{proof}
We prove the following by induction on $k$: if $L$ is
accepted by a DFA of size $(\lambda,\mu)$, where $n = \mu + \lambda$,
then for all $k \geq 2$, there exists a DFA $M$ of size
$(\lambda,k\mu + (k-1)\lambda - k + 1)$ such that $L(M) = L^k$.

If $k = 2$, then an easy application of Theorem~\ref{pig_sha}
with $L_1 = L_2 = L$ gives a DFA $M$ of size $(\lambda,2\mu + \lambda - 1)$
such that $L(M) = L^2$.

If $k > 2$, then write $L^k = L^{k-1}L$.  By induction, $L^{k-1}$
is accepted by a DFA of size $(\lambda,(k-1)\mu + (k-2)\lambda - k + 2)$.
Applying Theorem~\ref{pig_sha} with $L_1 = L^{k-1}$ and $L_2 = L$
gives a DFA $M$ of size $(\lambda,k\mu + (k-1)\lambda - k + 1)$ such
that $L(M) = L^k$.  The DFA $M$ thus has
\begin{eqnarray*}
&   & \lambda + k\mu + (k-1)\lambda - k + 1 \\
& = & k\mu + k\lambda - k + 1 \\
& = & k(\mu + \lambda) - k + 1 \\
& = & kn - k + 1
\end{eqnarray*}
states, as required.
\end{proof}

The following theorem gives a matching lower bound for the state
complexity of $L^k$.

\begin{theorem}
For any integers $n,k$, $n \geq 2$, $k \geq 2$, there exists a DFA
$M$ with $n$ states over a unary alphabet such that the minimal DFA
accepting the language $L^k(M)$ has $kn-k+1$ states.
\end{theorem}

\begin{proof}
We define a DFA $M = (Q,\Sigma,\delta,0,F)$, where
$Q = \lbrace 0,\ldots,n-1 \rbrace$, $\Sigma = \lbrace 0 \rbrace$,
$F = \lbrace n-1 \rbrace$, and for any $i$, $0 \leq i \leq n-1$,
$\delta(i,0) = i+1 \bmod n$.  The transition graph of $M$ is thus
a directed $n$-cycle.  Furthermore, $L(M) = 0^{n-1}(0^n)^*$.
Hence, $L^k(M) = (0^{n-1}(0^n)^*)^k = 0^{k(n-1)}(0^n)^*$.
The language $L^k(M)$ is accepted by the DFA
$M' = (Q',\Sigma,\delta',0,F')$, where
$Q' = \lbrace 0,\ldots,kn-k \rbrace$,
$F' = \lbrace kn-k \rbrace$, for any $i$, $0 \leq i < kn-k$,
$\delta'(i,0) = i+1$, and $\delta'(kn-k,0) = kn-k-n+1$. The DFA
$M'$ is minimal, since every unary accessible and
co-accessible DFA with a single final state is minimal.
\end{proof}

\section{Further work}
It remains to investigate the worst-case state complexity of
$L^3$, $L^4$, \emph{etc}.\ for general alphabets.

\section{Acknowledgements}
The inspiration to study this problem came from a talk given
by Sheng Yu at the ICALP 2004 Formal Languages Workshop (Colloquium in
honour of Arto Salomaa) that raised many interesting questions regarding
state complexity.  Thanks also to Jeffrey Shallit for suggesting this problem
and for helpful discussions along the way.


\begin{thebibliography}{99}
\bibitem{Bir92}
J.-C. Birget, Intersection and union of regular languages and
state complexity, Inform. Process. Lett. 43 (1992) 185--190.
\bibitem{Bir96}
J.-C. Birget, The state complexity of $\overline{\Sigma^*\overline{L}}$
and its connection with temporal logic, Inform. Process. Lett. 58 (1996)
185--188.
\bibitem{Chr86}
M. Chrobak, Finite automata and unary languages, Theoret. Comput. Sci.
47 (1986) 149--158.
\bibitem{HU79}
J.E. Hopcroft, J.D. Ullman, Introduction to Automata Theory,
Languages, and Computation, Addison-Wesley, 1979.
\bibitem{Jir05}
G. Jir\'askov\'a, State complexity of some operations on binary regular
languages, Theoret. Comput. Sci. 330 (2005) 287--298.
\bibitem{JJS04}
J. Jir\'asek, G. Jir\'askov\'a, A. Szabari, State complexity of
concatenation and complementation of regular languages, in:
Proc. 9th Internat. Conf. on Implementation and Application of
Automata (CIAA 2004), in: Lecture Notes in Comput. Sci., Vol. 3317,
Springer-Verlag, Berlin, 2004, pp. 178--189.
\bibitem{Ner58}
A. Nerode, Linear automaton transformations, Proc. Amer. Math. Soc. 9 (1958)
541--544.
\bibitem{PS02}
G. Pighizzini, J. Shallit, Unary language operations, state complexity
and Jacobsthal's function, Internat. J. Found. Comput. Sci. 13 (2002) 145--159.
\bibitem{RS59}
M.O. Rabin, D. Scott, Finite automata and their decision properties,
IBM J. Res. Develop. 3 (1959) 114--125.
\bibitem{YZS94}
S. Yu, Q. Zhuang, K. Salomaa, The state complexities of some basic
operations on regular languages, Theoret. Comput. Sci. 125 (1994) 315--328.
\end{thebibliography}
\end{document}